\documentclass[]{llncs}

\usepackage{geometry}
\usepackage[utf8]{inputenc}
\usepackage{caption}
\usepackage{amsfonts}
\usepackage{subcaption}
\usepackage{color}
\usepackage{booktabs}
\usepackage[table,xcdraw]{xcolor}
\usepackage{algorithm}
\usepackage{algpseudocode}
\usepackage{graphicx}
\usepackage[htt]{hyphenat}
\usepackage{listings}
\usepackage[normalem]{ulem}
\usepackage{amsmath}
\usepackage{multirow}
\usepackage{multicol}
\usepackage{booktabs}
\definecolor{dkgreen}{rgb}{0,0.6,0}
\definecolor{gray}{rgb}{0.5,0.5,0.5}
\definecolor{mauve}{rgb}{0.58,0,0.82}

\newcommand{\nand}{\texttt{NAND }}

\title{BitVMX: A CPU for Universal Computation on Bitcoin}

\author{Sergio Demian Lerner\inst{1,2}
\and Ramon Amela\inst{1}
\and Shreemoy Mishra\inst{1}
\and Martin Jonas\inst{2}
\and Javier {\'Alvarez Cid-Fuentes}\inst{1}
}

\institute{RootstockLabs \\ \email{\{sergio,ramon.amela,shreemoy,javier\}@rootstocklabs.com}
\and FairGate Labs \\ \email{\{sergio,martin.jonas\}@fairgate.io}}

\begin{document}
\raggedbottom

\pagestyle{plain}

\maketitle
\begin{abstract}
BitVMX is a new design for a virtual CPU to optimistically execute arbitrary programs on Bitcoin based on a challenge response game introduced in BitVM. 
Similar to BitVM1 we create a general-purpose CPU to be verified in Bitcoin script. Our design supports common architectures, such as RISC-V or MIPS. 
Our main contribution to the state of the art is a design that uses hash chains of program traces, memory mapped registers, and a new challenge-response protocol. We present a new message linking protocol as a means to allow authenticated communication between the participants. This protocol emulates stateful smart contracts by sharing state between transactions. This provides a basis for our verification game which uses a graph of pre-signed transactions to support challenge-response interactions.
In case of a dispute, the hash chain of program trace is used with selective pre-signed transactions to locate (via \textit{n-ary} search) and then recover the precise nature of errors in the computation. Unlike BitVM1, our approach does not require the creation of Merkle trees for CPU instructions or memory words. Additionally, it does not rely on signature equivocations. 
These differences help avoid complexities associated with BitVM1 and make BitVMX a compelling alternative to BitVM2. Our approach is quite flexible, BitVMX can be instantiated to balance transaction cost vs round complexity, prover cost vs verifier cost, and precomputations vs round complexity.
\end{abstract}

\section{Introduction}
\label{intro}

BitVMX is a new design for a simple and efficient CPU that can optimistically verify the execution of arbitrary programs on Bitcoin. TrueBit~\cite{teutsch17,teutsch23} introduced the \textit{verification game} paradigm to blockchain systems. In this paradigm, if all parties agree on the result of an off-chain computation, then no computation is performed on-chain. In case of a dispute, an on-chain interaction formally solves the dispute through a challenge-response type verification game.

This paradigm gained mainstream awareness with the introduction of \textit{fraud proofs} and \textit{optimistic rollups}. More recently, we have seen the emergence of rollups using a different paradigm relying on \textit{validation proofs} created using \textit{Zero Knowledge} cryptography. 

Such scaling solutions have been harder to build on top of Bitcoin. Some reasons include Bitcoin's restricted scripting language, the lack of features related to transaction introspection, and a restricted execution environment that does not offer any native method to persist state across transactions. This is where a recent proposal by Robin Linus, BitVM~\cite{linus23} made an important contribution by presenting the first attempt to bring the Disputable Computation paradigm to Bitcoin. Remarkably, Linus' proposal does not require any consensus changes or soft forks on Bitcoin. This paradigm is now a rapidly growing area of research. Most of the community interest is focused on two primary use cases: (i) building trust-minimized bridges for Bitcoin sidechains and rollups and (ii) optimistic verification of ZK proofs for rollups and other applications.

The original BitVM design was a theoretical proposition with restricted applicability. Later versions became more practical. BitVMX offers an alternative to these newer designs. BitVMX allows funds to be locked in a UTXO  with a spending constraint that depends on the result of running a predefined program with a given input. In its simplest form, BitVMX is  a two-party protocol where the first participant is the \textit{prover} (also called the\textit{ operator}) and the counterparty is called the \textit{verifier}. BitVMX can be extended to the $N$ verifier setting with the assumption that at least one of the $N\ge1$ verifiers is honest (as in BitVM).

When the operator wants to access the funds locked in the UTXO, he claims that he has executed the program correctly for some input of mutual interest, and shares that input  with the verifier (off-chain). The operator first claims that the computation yields an approval for the spending of the UTXO. If the verifier does not contest this claim, then the operator can access the funds after a timelock. The timelock allows the verifier to execute the program locally with the reported input and verify the operator's claim. If the verifier detects cheating, she can challenge the operator and both enter a dispute protocol (verification game) on the Bitcoin blockchain.

BitVMX improves BitVM by simplifying the way computations are represented. The protocol requires both parties to run the program locally. When executing the program, each party simultaneously generates an \textit{execution trace} and a \textit{hash chain}, with one hash for each step. The hash chain is constructed by (i) appending the current step's trace to the previous step's hash, and then (ii) hashing the result securely. Starting from an initial state, each step's hash is computationally unforgeable and part of a \textit{recursive hash chain}. Any difference in the program's execution trace across the parties will lead to diverging hash chains.

%The execution trace can either be saved or recomputed at every round at low cost.
The trace representation allows BitVMX to perform \textit{n-ary} searches, rather than just binary ones as in BitVM. This can significantly reduce the number of challenges required to find the conflicting computational step. Once a fault has been localized, a variety of specially crafted challenges can be used to determine the exact nature of the fault. For instance, BitVMX uses a new challenge-response protocol to detect faulty memory reads based on tracking the last time a memory word was written to.

All responses provided by the responder are uniquely linked to a specific challenge. The challenger can use the responder’s own response at later stages of the protocol to prove cheating and ultimately force a dishonest operator to concede the funds. Using the responder's own signed message helps BitVMX avoid the use of on-chain transactions to prove equivocations, which results in a simpler protocol. Equivocations can be used, as in BitVM, they are simply not core to BitVMX.
%\textcolor{red}{The rest of this paper is structuted as follows: ... (add this?) .. No. Don't bother.} 

\section{Related work}
\label{sec:related}

TrueBit was the first blockchain system to propose a verification game to validate complex off-chain computations without running them on-chain. In TrueBit, a \textit{solver} runs an arbitrary computation off-chain and posts the result on a smart contract for validation. The validation is performed through a verification game triggered by a challenger. As part of this game, the challenger forces the solver to reveal certain details of the execution in order to find a conflicting computational step.  In case of equivocation, the smart contract executes the conflicting step to penalize the solver. This reduces the amount of computation that has to be executed on-chain to a single step and extends the computational capabilities of the underlying blockchain. 

TrueBit's base concept was popularized by optimistic rollups on Ethereum. However, there was no such implementations on Bitcoin - until BitVM~\cite{linus23}. The first version of BitVM proposed a mechanism to create logic gate commitments in Bitcoin that enable the verification of arbitrary computations in a way similar to TrueBit. This is based on the fact that any computation can be expressed as a circuit of \nand gates. In BitVM, a prover and a verifier create a Taproot tree where each leaf represents one \nand gate. The two parties then pre-sign a sequence of challenge-response transactions that allow the verifier to challenge the inputs and output of any \nand gate in the circuit. The prover is forced to respond to the challenges by revealing hash preimages for the gate's inputs and output.  When challenged, gates are executed and verified on-chain. After a series of challenges, the verifier can punish the (dishonest) prover by forcing the publication of incorrect or conflicting data on-chain.

Though conceptually interesting, working directly with logic gates is not a practical approach for verifying complex computations that must ultimately be run using Bitcoin script. Linus~\cite{bitvm-cpu} presents an improved version, BitVM1, where computations are represented as a sequence of complex instructions of a virtual CPU. In BitVM1 each instruction consists of two inputs, an opcode, and an output. The state at each computational step is defined by the root of a Merkle tree that stores the contents of all the memory locations. A trace Merkle tree is built upon leaves that contain the memory states. The verification game consists of the verifier first identifying a conflicting computational step by binary search (or \textit{bisection}) over the trace Merkle tree, and then showing that the prover committed incorrect information on-chain. To detect incorrect memory reads or writes, an additional binary search is performed over the hashes of a Merkle path in the memory Merkle tree related to the step in question. As in TrueBit, at the end of the process, the system executes at most a single instruction on-chain. Unlike BitVM1, BitVMX does not use Merkle trees to store the computational trace, the program, or the memory. Our use of hash chains for the trace allows BitVMX to perform \textit{n-ary} searches. BitVM1 needs to partially or totally recompute the memory Merkle tree to reflect all memory updates performed at each step. BitVMX does not use trees, but it marks each memory word with an integer indicating the last step it was updated. This is more memory-efficient and also reduces computation costs.

Recently, the development of BitVM1 was paused in favor of a new approach in BitVM2~\cite{bitvm2}. In BitVM2, the full computation to be verified is divided into a sequence of connected steps representing intermediate computations. The prover commits to the inputs, the final result of the entire computation, and to all of the intermediate results in a single transaction. After this, \textit{any arbitrary} party can penalize the prover for submitting an incorrect intermediate result by forcing the on-chain execution of the particular step re-generating the result. BitVM2 allows anyone to become the verifier, which is a significant security improvement. BitVM2 also reduces the number of transactions needed in the verification game. However, this approach lacks the generality afforded by BitVMX (and BitVM1). For every program, each intermediate computational step will require a custom implementation in Bitcoin script. To avoid having to write custom implementations for each program, BitVM2 proposes writing a SNARK~\cite{bitansky12} verifier in Bitcoin script. This will enable the verification of any computation -- for which a SNARK proof can be generated. BitVMX, by contrast, is designed from the start as a general-purpose framework that enables verifying any program that can be compiled to common architectures, such as RISC-V, MIPS or any standard CPU.

\section{Message linking scheme}
\label{sec:linking}

Message linking is a simple cryptographic scheme used in BitVMX to link data referenced in sequential Bitcoin transactions. If a computation is disputed, then our verification protocol plays out on-chain over challenge and response transactions, the details of which appear in Section \ref{sec:crp}. We want these back and forth transactions to transfer some information about the state of the computation we wish to verify. Bitcoin has no native mechanism to transfer state data across transactions. So we present a framework to share state data - with a cryptographic link between a response message and the corresponding challenge message.% in a unique and provable manner.

We also want to ensure that if an operator is challenged by a verifier, then the operator must respond, and do so in a manner that is consistent with previously agreed upon terms.  The verifier's challenges must also fit a pre-arranged format. Implementing such restrictions can be challenging because Bitcoin script is not Turing complete. The execution environment has no access to transaction data.  So a coin's spending restrictions are typically limited to checking signatures, hash locks and time locks. There is no op-code for concatenation and arithmetic operations are limited to words of at most 32 bits (4 bytes). One consequence of these limitations is the current impossibility of \textit{stateful smart contracts} or \textit{covenants} on Bitcoin.

Some of the above restrictions on introducing state into transactions can be softened by all parties pre-signing a set of related transactions that form a directed acyclic graph. The parties can commit to all inputs and outputs as desired, and then a subset of these pre-signed transactions can be broadcast in some sequence to reproduce covenant-like behavior. To make such configurations robust, every party has to be sure that the other parties have no way to spend the locked funds in a way that can break the intended flow. This is related to the concept of \textit{connectors} introduced in the Ark protocol~\cite{ark}. BitVMX uses pre-signed transactions to use connectors and force the operator to engage in the message linking protocol. %The key idea connectors is to tie the spending conditions of a coin created in one transaction dependent on (the output of) another transaction.

Our message linking scheme is composed as follows. Both parties commit to the state data they intend to communicate by using \textit{one-time signature} schemes such as  Lamport~\cite{lamport79} or Winternitz~\cite{merkle89} signatures. We use transaction templates to specify the format of challenge and response transactions. Since enforcing a predefined sequence of transaction templates is not possible using Bitcoin script, the parties must jointly pre-sign the transactions  generated from templates during a \textit{setup phase}. When pre-signing, they must use appropriate \texttt{SIGHASH} flags to commit to all inputs and outputs - leaving only the witness data to be filled in later. This way, the transaction \textit{id}s will be fixed from the start and we can create sequential transactions with dependencies, so state information can be communicated across them.

An on-chain verification game can begin only after the operator broadcasts a \textit{claim} or \textit{kickoff} transaction to initiate a withdrawal of funds from the locked deposits. The actual challenge-response state information to be exchanged in the process of the verification game will obviously not be available during the setup phase. This implies that the commitment of state data using one-time signatures must happen later, by adding them to the segregated witness stack of appropriate pre-signed transactions (created from templates). It is only after the valid state data is added to the witness stack that the pre-signed templates become valid transactions that can be broadcast.

More formally, to create a link between two messages, a challenger $C$ and a responder $R$ start by pre-signing three transactions together: a start transaction $T_s$, and two segregated witness transaction templates $T_c$ and $T_r$ that do not include their witnesses. Additionally, $C$ and $R$ generate one-time signatures with public keys $P^o_c$ and $P^o_r$ respectively.

\begin{figure*}[t]
	\centering
	\includegraphics[width=\textwidth]{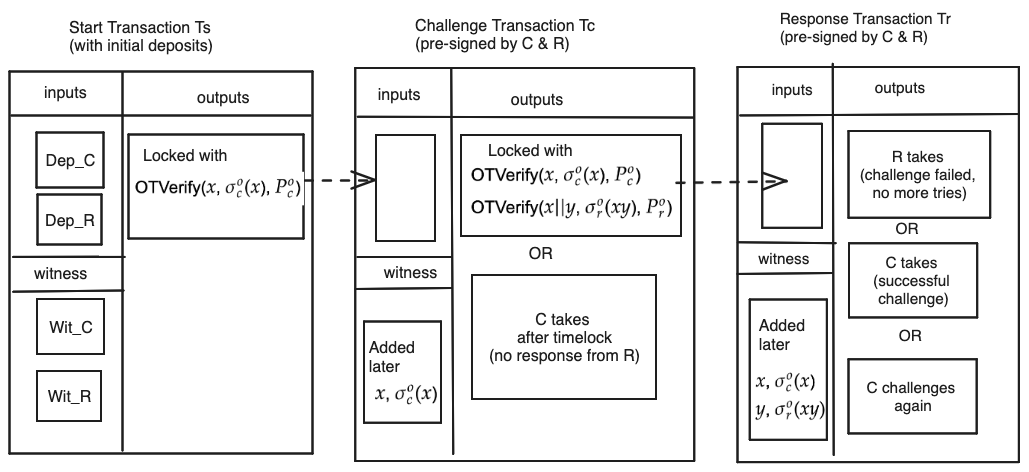}	
	\caption{Simplified illustration of using one-time signatures and pre-signed templates to link messages across transactions}
	\label{fig:linking:exacli}
\end{figure*}

The starting transaction $T_s$ contains an output that can be spent by $C$ by providing an arbitrary message $x$ along with a one-time signature $\sigma^o_c(x)$ that is consistent with $C$'s one-time public key $P^o_c$. Figure~\ref{fig:linking:exacli} shows a diagram of the linking scheme. To start the message linking scheme $C$ adds a message $x$ and a one-time signature $\sigma^o_c(x)$ to the witness stack of transaction $T_c$. The function \texttt{OTVerify} in Figure~\ref{fig:linking:exacli} verifies a one-time signature against a pre-defined (one-time) public key. \texttt{OTVerify} is an example of a \textit{meta} op-code. This is something that we must implement using Bitcoin script.

$T_c$ creates an output that can be spent by the responder $R$. In order to spend this output, $R$ must duplicate $C$'s message and signature ($x$, $\sigma^o_c(x)$) and add them to the witness along with a message $y$ and one-time signature  $\sigma^o_r(xy)$ of their own. $R$ actually signs $y$ after appending it to $x$. Thus, $R$'s signature $\sigma^o_r(xy)$ is over the concatenation of both messages $(x || y)$ and this must be consistent with $R$'s one-time public key $P^o_r$. Throughout this paper, $||$ denotes the concatenation operator.

Note that $C$ can use their private key to rebroadcast a different value $x'$ in the future. To avoid this, we require that if $C$ has to refer to their own message $x$ in some future transaction, then they must use the value from $R$'s response which contains both $x$ and $y$. Since only $C$ knows their one-time \textit{private} key, it is safe for anyone else (including $R$) to rebroadcast $C$'s message $x$.  

% \begin{figure*}[h]
% 	\centering
% 	\includegraphics[width=\textwidth]{img/linking.png}	
% 	\caption{}
% 	\label{fig:linking}
% \end{figure*}

Note that $C$ is forced to sign and reveal $x$ in order to start the challenge and that $R$ is forced to repeat the provided $x$ and respond with their own signed message. $T_c$ contains another output that pays to $C$ if $R$ does not publish $T_r$ before a deadline, which forces $R$ to sign and reveal $y$ once $T_c$ is published. The sequence of transactions can be extended to an arbitrary length by replicating the scheme after $T_r$. %At any point after $T_r$, $C$ can use $x$ and $y$ as proof that $R$ provided incorrect data. Neither $C$ nor $R$ can use other values for $x$ and $y$ than the values committed in $T_c$ and $T_r$.

Figure \ref{fig:linking:exacli} is a highly simplified illustration of how message linking can be used in BitVMX. An actual implementation of the protocol will be more complex - because message linking is only a part of the challenge-response protocol. For instance, the initial deposits from each party will typically be locked in a transaction that is mined much earlier than shown here. Once the deposits are locked, all parties are committed to the protocol.  The full protocol, presented later, will require pre-signing a set of transactions that form a directed acyclic graph. It is possible to pre-sign several transactions with the same input but different outputs. By doing so, we can create a decision tree where the party in charge of publishing a certain message at a particular step of the protocol chooses the branch through which the protocol should continue. When one path is chosen, some others get discarded as UTXOs associated with the alternative paths have been spent. The initial transaction generation, pre-signing and sharing process is called a \textit{setup ceremony}.

\section{CPU specification}
\label{sec:specification}

BitVMX defines a virtual CPU to represent the computation that needs to be verified. This virtual CPU is specified by a set of instructions, addressable memory, and a representation of the CPU state. BitVMX provides a flexible design that allows the implementation of common architectures, from 8-bit CPUs to 32-bit RISC-V, without having to build a specific compiler for BitVMX.

A verifiable program in BitVMX is a sequence of executed instructions or steps in this virtual CPU, where each step is structured as follows:

\begingroup
\setlength{\tabcolsep}{1.7pt}
\begin{table}[h!]
\begin{tabular}{|cccccccc|ccc|}
\hline
\multicolumn{8}{|c|}{\textbf{Inputs}}                                                                                                                                                                                                                                                                  & \multicolumn{3}{c|}{\textbf{Outputs}}                                                          \\ \hline
\multicolumn{3}{|c|}{\texttt{read1}}                                                                                  & \multicolumn{3}{c|}{\texttt{read2}}                                                                                  & \multicolumn{2}{c|}{\texttt{readPC}}                    & \multicolumn{2}{c|}{\texttt{write}}                                        & \texttt{writePC} \\ \hline
\multicolumn{1}{|c|}{\texttt{address}} & \multicolumn{1}{c|}{\texttt{value}} & \multicolumn{1}{c|}{\texttt{lastStep}} & \multicolumn{1}{c|}{\texttt{address}} & \multicolumn{1}{c|}{\texttt{value}} & \multicolumn{1}{c|}{\texttt{lastStep}} & \multicolumn{1}{c|}{\texttt{address}} & \texttt{opcode} & \multicolumn{1}{c|}{\texttt{address}} & \multicolumn{1}{c|}{\texttt{value}} & \texttt{pc} \\ \hline
\end{tabular}
\caption{Layout of an instruction executed by the CPU.}
\label{tab:instructions}
\end{table}
\endgroup

Every executed instruction reads at most two values and an op-code from memory, performs an operation, and writes the result to memory. Additionally, each instruction sets the address of the next instruction by modifying the program counter (\texttt{pc}). The field named \texttt{lastStep} stores the program step number where each of the read values was last written. Using this instruction structure, we provide the following definitions:

\begin{definition}
    \label{def:program}
    \normalfont
    We define the \texttt{full\_trace} of an execution step $i$ as the concatenation of all the input and output fields in Table~\ref{tab:instructions} 
    \begin{equation*}
        \texttt{full\_trace}_i = \texttt{read1.address}_i || \texttt{read1.value}_i || \texttt{read1.lastStep}_i || ... || \texttt{writePC.pc}_i   
    \end{equation*}    
\end{definition}

\begin{definition}
    \label{def:trace}
    \normalfont
    We define the \texttt{trace} of an execution step $i$ as the concatenation of data from just the output fields, that is,
    \begin{equation*}
        \texttt{trace}_i = \texttt{write.address}_i || \texttt{write.value}_i || \texttt{writePC.pc}_i    
    \end{equation*}    
\end{definition}

The length of the memory addresses depends on the architecture. For example, they are 32-bit long for a 32-bit RISC-V. Registers are mapped to pre-defined memory addresses. For example, we can map $r$ registers to the memory addresses from $0$ to $r-1$, and the program counter is the only register that does not have a memory address.

The state of the virtual CPU consists of the values in every memory address at each program step plus the \texttt{pc}. We represent \textit{state transitions} using a hash chain where each member, denoted by \texttt{stepHash}$_i$, is constructed recursively as follows.

\begin{definition}
    \label{def:stephash}
    \normalfont
     Given a hash function $h$, we define \texttt{stepHash} for an instruction $i$ as
     \begin{align*}
         \texttt{stepHash}_i = h(\texttt{stepHash}_{i-1} || \texttt{trace}_i) .
    \end{align*}    
\end{definition}

With this recursive construction, each \texttt{stepHash} implicitly depends on every memory write operation from the start to the current instruction step. By using a secure hashing function, we can be almost certain that each \texttt{stepHash} is a unique representation of the memory state at each program step.

Arithmetic operations in Bitcoin script take 32-bit signed inputs. Thus, implementing 32 or 64-bit architectures requires a mechanism to handle data in smaller chunks using Bitcoin primitives. This means we cannot use the built-in hash functions available in Bitcoin script to compute the hash of the above objects. The chosen hash function, $h$,  has to be implemented using basic arithmetic and logical primitives of Bitcoin script. Therefore, both the choice of a hash function and its implementation are important for efficiency and security of the protocol.

To initialize the CPU, we insert write instructions at \textit{negative} \texttt{full\_trace} positions that load the program into memory. That is, if a program to be executed has $m$ instructions, then \texttt{full\_trace}$_{-m}$ to \texttt{full\_trace}$_{-1}$ contain only write operations for each instruction. 

We set the starting point of the hash chain as
\begin{equation*}
    \texttt{stepHash}_{-m-1} = 0
\end{equation*}
and then use Definition~\ref{def:stephash} to compute \texttt{stepHash}$_{i}$ for all steps from $i=-m$ up to $i=-1$. In this way, \texttt{stepHash}$_{-1}$ represents the initial state of the CPU, where each program instruction is in a particular memory address. The range of memory addresses that store the program is read-only. Additionally, we define another range of read-only memory addresses to store the program input. The rest of the memory addresses are initialized with the value 0 and their \texttt{lastStep} field is initialized with the special value \texttt{INITIAL}. Given some input, we can execute the program. If the program exits successfully at step $i=k$, then we set \texttt{stepHash}$_{i} = 1$ for all $i>k$. In case of an exception, we set these values to a negative number.

\section{Challenge-response protocol}
\label{sec:crp}

In this section, we describe the challenge-response protocol that BitVMX uses to verify computations on-chain. The protocol is based on the message linking scheme described in Section~\ref{sec:linking} and involves a prover and a verifier. 

During setup, the prover and the verifier agree on \texttt{stepHash}$_{-1}$, and before running the protocol both parties initialize the CPU as described in Section~\ref{sec:specification}. After this, the prover can trigger the challenge-response process by providing the input of the program and the \texttt{stepHash} value of the very last step. The protocol continues if the verifier disagrees with the final state of the CPU for the given input. In that case, the goal of the verifier is to identify an error in the prover's computation. Towards this, the verifier starts by identifying the first conflicting computational step and then forcing the prover to reveal the data associated with that step. Then, the verifier identifies the precise nature of the error in this step and validates the error on-chain. In the following sections, we describe various parts of the challenge-response process in more detail.

\begin{figure}[b!]
	\centering
	\includegraphics[width=0.5\textwidth]{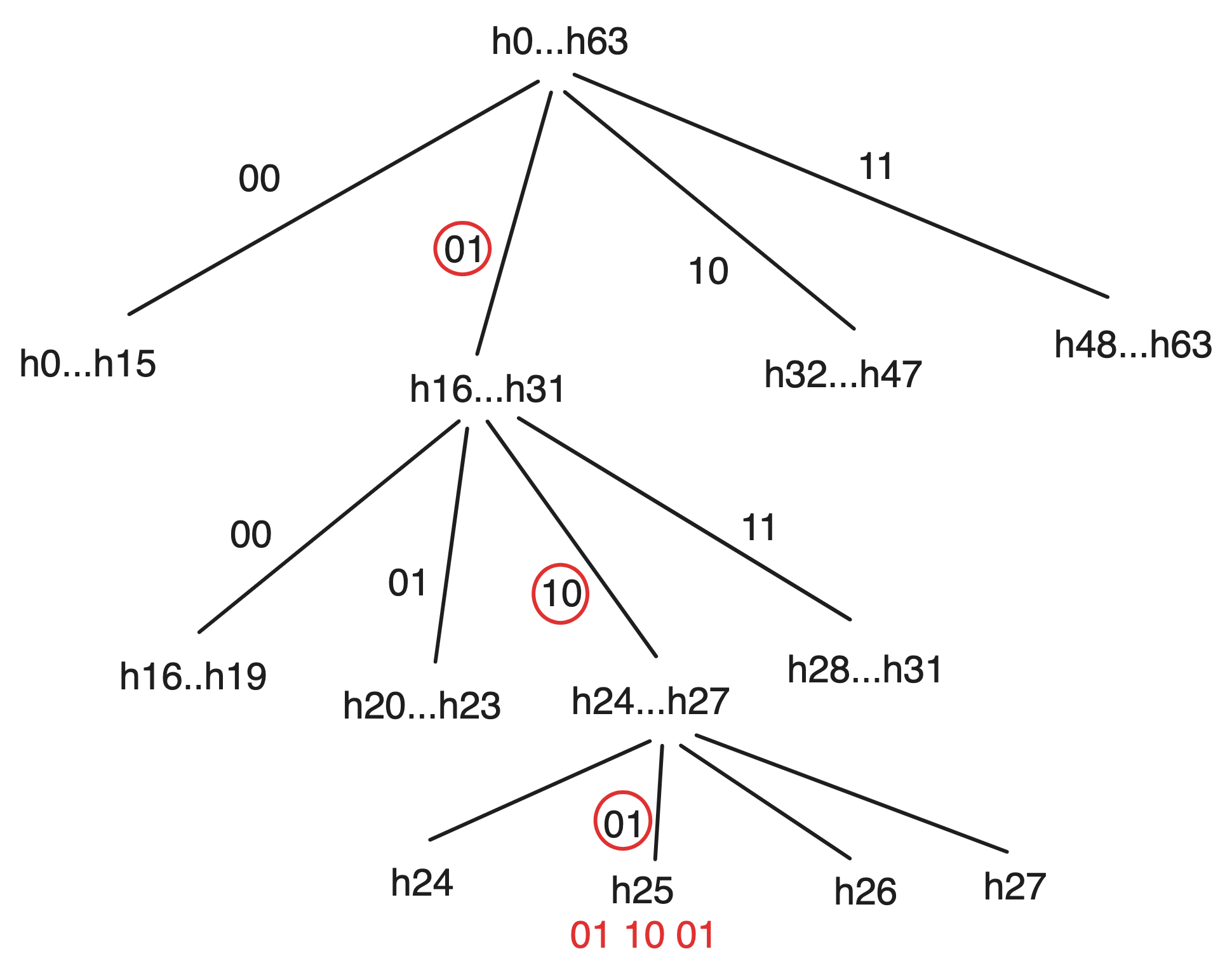}	
	\caption{\textit{n-ary} search mechanism with 64 hashes ($m=64$, $n=4$, $l=3$). The concatenation of the interval identifiers selected produces $011001_b = 25$.}
	\label{fig:search}
\end{figure}

\subsection{Search mechanism}
\label{sec:search}

At various parts of the challenge-response protocol, the prover and the verifier use an \textit{n-ary} search algorithm. Given a sequence of \texttt{stepHash} values, we split the search space into $n$ intervals defined by $n-1$ equally spaced hashes, where $n$ is a power of two. We identify each interval with the \textit{binary representation} of a number in the range $0$ to $n-1$. We then chose one of the intervals and split it into a further $n$ sub-intervals. After $\log_{n}(m)$ such splits, let us assume we reach the $i$th hash in the sequence. Suppose we choose $m$ such that $m = n^l$ for some fixed number of iterations $l$. Then the concatenation of the identifier for each interval selected during the search provides the index (in binary) of the $i$-th hash. Figure~\ref{fig:search} illustrates an \textit{n-ary} search process that reaches the 25th hash in the sequence after $l=3$ iterations with $m=64$ and $n=4$.

Note that every interval of the search process can be defined with $n-1$ hashes, because both left and right boundaries are provided in the previous iteration. For example, in Figure~\ref{fig:search}, the interval $h_{16}, ..., h_{31}$ can be defined by $h_{19},h_{23},h_{27}$.

\subsection{Conflicting step search}
\label{sec:step-search}

A challenge is needed only when the two parties disagree on the last \texttt{stepHash} reported by the prover. Since the prover and the verifier agree on \texttt{stepHash}$_{-1}$ during setup and disagree on the last one, there must be two consecutive hashes \texttt{stepHash$_i$} and \texttt{stepHash}$_j$, such that \texttt{stepHash}$_i$ is correct and \texttt{stepHash}$_j$ is incorrect according to the verifier. The goal of this stage of the protocol is to identify a pair $i,j$ for which this is true. We refer to \texttt{stepHash}$_j$ as the first conflicting step. 

\begin{figure}[b!]
	\centering
	\includegraphics[width=\textwidth]{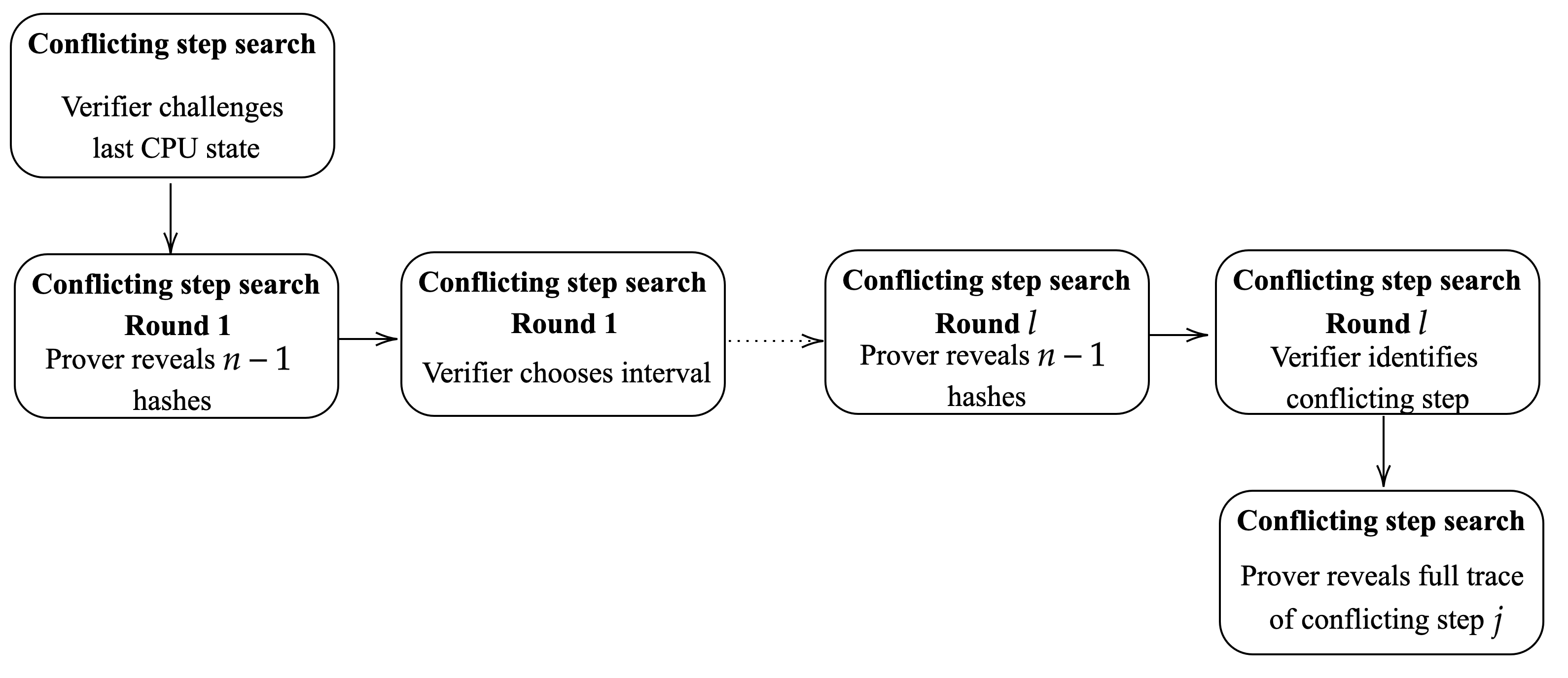}	
	\caption{First stage of the challenge-response protocol used to identify the first conflicting step in the sequence of \texttt{stepHash} hashes.}
	\label{fig:step-search}
\end{figure}

Figure~\ref{fig:step-search} illustrates the first stage of the challenge-response protocol, which begins when the verifier challenges the final state of the CPU published by the prover. To identify the first conflicting step, the verifier runs an \textit{n-ary} iterative search with the prover for a number of rounds as described in Section~\ref{sec:search}. In this process, the prover reveals $n-1$ hashes and the verifier responds by choosing the lowest interval defined by $i,j$, such that \texttt{stepHash}$_i$ is correct and \texttt{stepHash}$_j$ is incorrect. In the following round, the prover reveals $n-1$ new hashes between the previous $i,j$ interval. After $l$ rounds, the verifier reaches consecutive values for $i,j$ and the first conflicting step in the chain is \texttt{stepHash}$_j$. 

Note that this \textit{n-ary} search process creates a trade-off between its duration and the amount of data published per round. For example, for $2^{32}$ computational steps, the protocol needs 32 rounds with $n=2$, 16 rounds with $n=4$, and 8 rounds with $n=16$. During the setup stage, the prover and the verifier decide on a maximum number of rounds $l$ for the search process, which limits the maximum number of program steps to $m=n^l$.

Once the verifier identifies the first conflicting step $j$, the prover is asked to reveal \texttt{full\_trace}$_j$, which contains all the information of the conflicting step and sets up the next challenge.

%\textcolor{blue}{In addition, it is important to limit the total amount of accepted steps as much as possible while ensuring that the desired computation can be done. \newline
%Furthermore, there is a final property that needs to be taken into account when choosing both parameters. By choosing M and N in such a way that N+1 is a power of 2, the concatenation of the interval indexes unveiled at each iteration gives as a result the final step number. Hence, the exact step number can be easily be computed with the information committed on-chain. Next, the prover unveils the whole specification of the step. Hence, at the end of this phase, both the prover and verifier agree on the first wrong stepHash and the full specification delivered by the prover.}

\subsection{Conflicting step challenge}

The second stage of the challenge-response protocol continues after the prover reveals \texttt{full\_trace}$_j$ with the full description of the first conflicting step $j$. In this stage, the verifier challenges specific data of the conflicting step by choosing one of several possible paths in the protocol. In case the verifier disagrees with multiple parts of the conflicting step, the rational choice is to challenge the part that requires the fewest interactions to verify. In the following, we describe each challenge type in detail.

\subsubsection{Trace hash challenge}

The first information about the conflicting step that the verifier can challenge is the \texttt{stepHash} value itself. At this point, the prover has revealed \texttt{full\_trace}$_j$ (including \texttt{trace}$_j$), \texttt{stepHash}$_j$ and \texttt{stepHash}$_{j-1}$ for the first conflicting step $j$. Therefore, the verifier can show that the prover is wrong by validating the following inequality on-chain.
\begin{equation*}
    \texttt{stepHash}_j \neq h(\texttt{stepHash}_{j-1}||\texttt{trace}_j)
\end{equation*}

\subsubsection{Program input challenge}

During setup, the prover and the verifier agree on a read-only memory range for storing the input of the program. The prover must provide such input in order to validate the computation. The verifier can challenge an input value in either \texttt{read1} or \texttt{read2} by showing on-chain that \texttt{read.address}$_j$ in an input memory address and that \texttt{read.value}$_j$ is different from the value provided by the prover for that input.

\subsubsection{Read value challenge}
\label{sec:laststep}

The verifier can challenge \texttt{read.value}$_j$ for either \texttt{read1} or \texttt{read2} by checking the \texttt{read.lastStep}$_j$ value provided by the prover. The verifier can easily validate on-chain that $0 \le \texttt{read.lastStep}_j \le j$, so we assume that the \texttt{lastStep} value is within boundaries.

The verifier requests different information from the prover depending on the value in \texttt{read.lastStep}$_j$ to show that \texttt{read.value}$_j$ is incorrect. If the prover reveals incorrect data during this challenge, the verifier runs the challenge described in Section~\ref{sec:lasthash}. There are three possible scenarios:
\begin{enumerate}
    \item The \texttt{lastStep} value is marked as \texttt{INITIAL} but the address has been written to in a previous instruction
    \item The \texttt{lastStep} value revealed is greater than or equal to the correct value
    \item The \texttt{lastStep} value revealed is lower than the correct value
\end{enumerate}

In the first scenario, the verifier requests \texttt{stepHash}$_k$ and \texttt{trace}$_k$ from the prover, such that
\begin{equation*}
    \texttt{write.address}_k = \texttt{read.address}_j ,
\end{equation*}
which proves on-chain that $\texttt{read.lastStep}_j \neq \texttt{INITIAL}$.

In the second scenario, the verifier requests \texttt{stepHash}$_k$ and \texttt{trace}$_k$ from the prover, such that $k=\texttt{read.lastStep}_j$ and shows on-chain that
\begin{gather*}
    \texttt{write.address}_k \neq \texttt{read.address}_j \\
    \textrm{or} \\
    \texttt{write.value}_k \neq \texttt{read.value}_j ,
\end{gather*}
which proves that either the instruction $k$ does not write $\texttt{read.address}_j$ or that the value written is not the value read.

In the third scenario, the verifier requests \texttt{stepHash}$_k$ and \texttt{trace}$_k$ from the prover, such that step $k$ is the actual last step that writes the value. This shows on-chain that
\begin{gather*}
    \texttt{write.address}_k = \texttt{read.address}_j \\
    k > \texttt{read.lastStep}_j , 
\end{gather*}
which proves that $\texttt{read.lastStep}_j$ is incorrect. 

If the prover reveals an incorrect \texttt{stepHash}$_k$ in any of the three scenarios, the verifier triggers a new \textit{n-ary} search to challenge the hash. We describe this process in Section~\ref{sec:lasthash}.

\subsubsection{Last step hash challenge}
\label{sec:lasthash}

The verifier triggers a new \textit{n-ary} search process on \texttt{stepHash} if the prover reveals an incorrect \texttt{stepHash}$_k$ as part of the read value challenge (Section~\ref{sec:laststep}). The goal of the verifier is to find $q$ such that \texttt{stepHash}$_{q-1}$ is incorrect and \texttt{stepHash}$_q$ is correct. This process is equivalent to the search process described in Section~\ref{sec:step-search} but with the correct and incorrect hashes reversed. The verifier is guaranteed to find such $q$ because the prover revealed the correct value for \texttt{stepHash}$_{j-1}$ during the search for the first conflicting step and now the prover has revealed an incorrect hash in step $k<j-1$. After identifying $q$, the verifier forces the prover to reveal \texttt{trace}$_q$ and shows on-chain that 
\begin{equation*}
    h(\texttt{stepHash}_{q-1} || \texttt{trace}_q) \neq \texttt{stepHash}_q
\end{equation*}
and thus, that \texttt{stepHash}$_{q-1}$ is incorrect.

\begin{figure*}[h]
	\centering
	\includegraphics[width=0.5\textwidth]{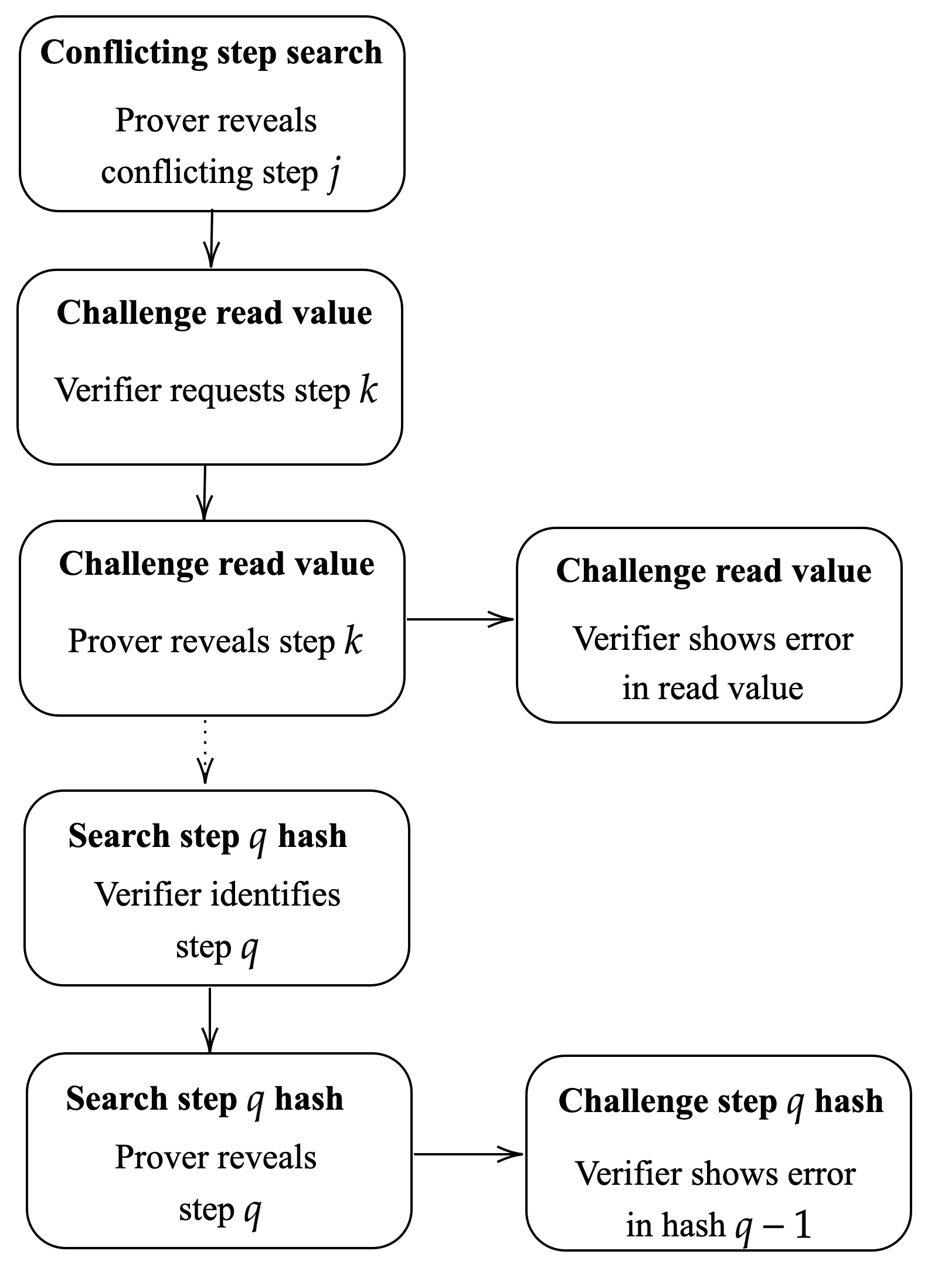}	
	\caption{Read value and last step hash challenges. See Figure~\ref{fig:step-search} for the detailed search process.}
	\label{fig:laststep}
\end{figure*}

Figure~\ref{fig:laststep} shows a diagram of the read value and last step hash challenges. We omit the rounds of the search process for simplicity. 

\subsubsection{Program counter challenge}

The verifier can show an error in the program counter of the first conflicting step, that is \texttt{readPC.address}$_j$, by providing \texttt{stepHash}$_{j - 2}$ and \texttt{trace}$_{j - 1}$, and validating on-chain that
\begin{align*}
    & h(\texttt{stepHash}_{j - 2} || \texttt{trace}_{j - 1}) = \texttt{stepHash}_{j - 1} \\
    & \texttt{writePC.address}_{j - 1} \neq \texttt{readPC.address}_j .
\end{align*}

The first validation guarantees that the verifier provides correct values for \texttt{stepHash}$_{j - 2}$ and \texttt{trace}$_{j - 1}$ by checking them against \texttt{stepHash}$_{j - 1}$, which the prover revealed in the first stage of the protocol. The second validation shows that \texttt{readPC.address}$_j$ is incorrect.

\subsubsection{Op-code challenge}

The verifier can challenge the value in \texttt{readPC.opcode}$_j$ by initiating a step search as described in Section~\ref{sec:lasthash} in the negative positions of \texttt{stepHash}. That is, from \texttt{stepHash}$_{-p-1}$ to \texttt{stepHash}$_{-1}$. Recall that, during setup, the prover and the verifier insert write operations for each instruction in \texttt{full\_trace}$_{(-p-1)...-1}$ and agree on \texttt{stepHash}$_{-1}$. This search process allows the verifier to force the prover to reveal \texttt{full\_trace}$_{-k}$, where $-k$ is the write operation for instruction $j$. The verifier can then show on-chain that the address where the instruction is written matches \texttt{readPC.address}$_j$ but that the op-code written is different from \texttt{readPC.opcode}$_j$.

\subsubsection{Execution challenge}

The verifier can challenge the execution of \texttt{full\_trace}$_j$ if either \texttt{write.value}$_j$ or \texttt{writePC.pc}$_j$ are incorrect. This is done by executing the instruction on-chain and validating both writes.

\section{Conclusion and future work}
\label{sec:conclusion}

BitVMX expands the current state of the art in optimistic execution of 
off-chain programs on Bitcoin -- an area of research pioneered by BitVM. In this paper we laid out the primitives of BitVMX, focusing on the single verifier case. Future work can explore ways to extend this to the $N$ verifier setting in a robust manner. Our goal is to create a secure, extensible, open-source, peer-reviewed and sidechain-agnostic framework that can be used to develop blockchain bridges, aggregator oracles, and SNARK/STARK verifiers.

Further research is also needed on the economic incentives for the core protocol and to match the needs of specific use cases. This includes the size of deposits from prover and verifier, bounties for verifiers, the cost of capital when operating a bridge, and the crypto-economic security of the protocol. Applications may have varying priorities with regards to transaction costs or round complexity. Such trade-offs depend on opportunity costs inherent to each application. Future research can explore how BitVMX can be customized to suit the needs of distinct applications.

\bibliographystyle{splncs04}
\bibliography{biblio}

\end{document}